\documentclass[preprint,aps,12pt,showpacs,nofootinbib,tightenlines]{revtex4}
\usepackage{amsmath}
\usepackage{amssymb}
\usepackage{epsfig}
\usepackage{graphicx}
\def\be{\begin{eqnarray}}
\def\en{\end{eqnarray}}
\def\non{\nonumber\\}
\def \zpc{  Z. Phys. C  }

\def\prd{{Phys. Rev. D}~}
\def\prl{{ Phys. Rev. Lett.}~}
\def\plb{{ Phys. Lett. B}~}

\newcommand{\ppi}{\phi_{\pi}^A}
\newcommand{\ppip}{\phi_{\pi}^P}
\newcommand{\ppit}{\phi_{\pi}^T}

\newcommand{\psl}{ P \hspace{-2.4truemm}/ }
\newcommand{\nsl}{ n \hspace{-2.2truemm}/ }
\newcommand{\vsl}{ v \hspace{-2.2truemm}/ }
\newcommand{\epsl}{\epsilon \hspace{-1.8truemm}/\,  }
\begin{document}
%%--------------------------------------------
\title{The $B\to D_s^{(*)}\pi$ decays in the perturbative QCD }
\author{ Zhi-Qing
Zhang\footnote{Electronic address: zhangzhiqing@haut.edu.cn}
} \affiliation{
{\it \small  Department of Physics, Henan University of Technology,
Zhengzhou, Henan 450052, P.R.China}
} %%
\begin{abstract}
\date{\today}

In this paper, we calculate the branching ratios for $B^0\to D_s^+\pi^-, B^+\to
D_s^+\pi^0$, $B^0\to D_s^{*+}\pi^-$ and $ B^+\to D_s^{*+}\pi^0$ decays in the perturbative QCD factorization approach. We find that
the calculated branching ratios of these four decay channels agree well with the measured values and current experimental upper limit.
In the numerical calculation, we take the decay constant  and the shape parameter of the vector meson $D^{*}_s$
as $f_{D^{*}_s}=312$ MeV and $a_{D^{*}_s}=0.78$ respectively, which are larger than those in the previous calculations.

%\keywords{pQCD approach; B meson; decay constant.}
\end{abstract}

\vspace{1cm}

\maketitle

%=======================================================================
%                     Introduction
%=======================================================================

\section{Introduction}\label{intro}
In recent years, more and more effort has been made to the B meson decays with one \cite{cdlv1} even two \cite{cdlv2}
charmed mesons in the final states and it is found that the perturbative QCD factorization (pQCD) approach does work well in these decays. We will
calculate the branching ratios for the $B\to D^{(*)}_s\pi$ decays, which are shown in figure~1, by employing the pQCD approach.
The momenta of the two outgoing mesons are both approximately $\frac{1}{2}m_B(1-m^{2}_{D^{(*)}_s}/m_B^2)$. This is still large enough to
make a hard intermediate gluon in the hard part calculation. Most of the momenta
come from the heavy b quark in quark level. The light quark u (d) inside $B^+$ $(B^0)$ meson, which is usually
called spectator quark, carries small momentum of order of $\Lambda_{QCD}$. In order to form a fast moving light meson, the
spectator quark need to connect the four-quark operator $(\bar b u)_{V-A}(\bar c s)_{V-A}$ through an energetic gluon. The
hard four-quark dynamic together with the spectator quark becomes six-quark effective interaction. Since six-quark interaction
is hard dynamics, it is perturbatively calculable in theory.

On the experimental side, the branching ratios of $B^0\to D_s^+\pi^-, B^+\to
D_s^+\pi^0$ and $B^0\to D_s^{*+}\pi^-$ have been measured by BaBar \cite{barbar1} and Belle \cite{belle}. For $B^+\to D_s^{*+}\pi^0$ decay, only the
experimental limit is given by CLEO \cite{cleo0}. We list their values in the following \cite{pdg08}:
\be
Br(B^0\to D_s^+\pi^-) &=& (1.53\pm0.35)\times 10^{-5},\non
Br(B^+\to D_s^+\pi^0) &=& (1.6\pm0.6)\times 10^{-5}, \non
Br(B^0\to D_s^{*+}\pi^-) &=& (3.0\pm0.7)\times 10^{-5},\non
Br(B^+\to D_s^{*+}\pi^0) &<& 2.7\times 10^{-4}.
\en

This paper is organized as follows. In Sect.\ref{proper}, the light-cone wave functions of the initial and
the final state mesons   are  discussed.
In Sec.\ref{results}, we calculate analytically the related Feynman diagrams and present the various decay amplitudes
for the studied decay modes. The numerical results and the discussions are given
in the section \ref{numer}. The conclusions are presented in the final part.

%=======================================================================
%       Physical properties of $f_0(980)$ and $f_0(1500)$
%=======================================================================

\section{Wave functions of initial and final state mesons}\label{proper}
In pQCD calculation, the light-cone wave functions are nonperturbative and not calculable, but they
are universal and channel independent for all the hadronic decays.

As a heavy meson, the B meson wave function is not well defined. In general, the B meson light-cone matrix element can be decomposed
as \cite{grozin}
\be
&&\int_0^1\frac{d^4z}{(2\pi)^4}e^{i\bf{k_1}\cdot z}
   \langle 0|\bar{b}_\alpha(0)d_\beta(z)|B(p_B)\rangle \nonumber\\
&=&-\frac{i}{\sqrt{2N_c}}\left\{(\psl_B+m_B)\gamma_5 \left[\phi_B
({\bf k_1})-\frac{\nsl-\vsl}{\sqrt{2}} \bar{\phi}_B({\bf
k_1})\right]\right\}_{\beta\alpha}, \label{aa1}
\en
where $n=(1,0,{\bf 0_T})$, and $v=(0,1,{\bf 0_T})$ are the
unit vectors pointing to the plus and minus directions,
respectively. Because the contribution of the second Lorentz structure $\bar{\phi}_B(x,b)$ is numerically small and can
be neglected, we only consider the
contribution of the Lorentz structure:
\be
\Phi_B(x,b)=
\frac{1}{\sqrt{2N_c}} (\psl_B +m_B) \gamma_5 \phi_B (x,b).
\label{bmeson}
\en
.

In the heavy quark limit, we take the wave functions for the pseudoscalar meson $D_s$ and the vector meson $D^*_s$ as
\be
\Phi_{D_S}(x,b)=
\frac{1}{\sqrt{2N_c}} \gamma_5(\psl_{D_s} +m_{D_s})  \phi_{D_s} (x,b),\\
\Phi_{D^*_s}(x,b)=
\frac{1}{\sqrt{2N_c}} \epsl(\psl_{D^*_s} +m_{D^*_s})  \phi_{D^*_s} (x,b),
\label{dsp}
\en
where the polar vector $\epsl=\frac{M_B}{\sqrt{2}M_{D^*_s}}(1,-r^2_{D^*_s},{\bf 0_T})$. In the considered decays,
the $D^*_s$ meson is longitudinally polarized, so we only need to consider its wave function
in longitudinal polarization.

The wave function for the light pseudoscalar meson $\pi$  is given as
\be
\Phi_{\pi}(P,x,\zeta)\equiv \frac{1}{\sqrt{2N_C}}\gamma_5 \left
[ \psl \phi_{\pi}^{A}(x)+m_0^{\pi} \phi_{\pi}^{P}(x)+\zeta m_0^{\pi}
(\vsl \nsl - v\cdot n)\phi_{\pi}^{T}(x)\right ] ,
\en
where $P$ and
$x$ are the momentum and the momentum fraction of $\pi$ meson,
respectively. The parameter $\zeta$ is either $+1$ or $-1$ depending
on the assignment of the momentum fraction $x$. The chiral scale parameter $m_0^{\pi}$ is defined as $m_0^{\pi}=m^2_{\pi}/(m_u+m_d)$.
%===========================================================================
%                    Decay amplitudes in PQCD approach
%============================================================================

\section{ the perturbative QCD  calculation} \label{results}

Using factorization theorem, we can separate the decay amplitude into soft, hard, and harder dynamics characterized
by different scales, conceptually expressed as the convolution,
\be
{\cal A}(B \to D^{(*)}_s\pi)\sim \int\!\! d^4k_1
d^4k_2 d^4k_3\ \mathrm{Tr} \left [ C(t) \Phi_B(k_1) \Phi_{D^{(*)}_s}(k_2)
\Phi_{\pi}(k_3) H(k_1,k_2,k_3, t) \right ], \label{eq:con1}
\en
where $k_i$'s are momenta of light anti-quarks included in each meson, and
$\mathrm{Tr}$ denotes the trace over Dirac and color indices. $C(t)$
is the Wilson coefficient which results from the radiative
corrections at a short distance. In the above convolution, $C(t)$
includes the harder dynamics at a larger scale than that at the $M_B$ scale and
describes the evolution of local $4$-Fermi operators from $m_W$ (the
$W$ boson mass) down to $t\sim\mathcal{O}(\sqrt{\bar{\Lambda} M_B})$
scale, where $\bar{\Lambda}\equiv M_B -m_b$. The function
$H(k_1,k_2,k_3,t)$ describes the four quark operator and the
spectator quark connected by
 a hard gluon whose $q^2$ is in the order
of $\bar{\Lambda} M_B$, and includes the
$\mathcal{O}(\sqrt{\bar{\Lambda} M_B})$ hard dynamics. Therefore,
this hard part $H$ can be perturbatively calculated. The function
$\Phi_{(D^{(*)}_s, \pi)}$ are the wave functions of $D^{(*)}_s$ and $\pi$.

Since the $b$ quark is rather heavy, we consider the B meson at rest for simplicity. It is convenient to use the light-cone
coordinate $(p^+, p^-, {\bf p}_T)$ is used to describe the meson's momenta:
\be
p^\pm =\frac{1}{\sqrt{2}} (p^0 \pm p^3), \quad {\rm and} \quad {\bf p}_T =(p^1, p^2).
\en
At the rest frame of $B$ meson, the light meson moves very fast and so $P_3^+$ or $P_3^-$ can be treated as zero.
Using these coordinates, the B meson and the two final state meson momenta can be written as
\be
P_B =\frac{M_B}{\sqrt{2}} (1,1,{\bf 0}_T), \quad P_{2} =
\frac{M_B}{\sqrt{2}}(1,r^2,{\bf 0}_T), \quad P_{3} =
\frac{M_B}{\sqrt{2}} (0,1-r^2,{\bf 0}_T),
\en
respectively, where $r=M_{D^{(*)}_s}/M_B$. Putting the light anti-quark momenta in $B$,
$D^{(*)}_s$ and $\pi$ mesons as $k_1$, $k_2$, and $k_3$, respectively, we can
choose
\be
k_1 = (x_1 P_1^+,0,{\bf k}_{1T}), \quad k_2 = (x_2
P_2^+,0,{\bf k}_{2T}), \quad k_3 = (0, x_3 P_3^-,{\bf k}_{3T}).
\en
For these considered decay channels, the integration over $k_1^-$,
$k_2^-$, and $k_3^+$ in equation (\ref{eq:con1}) will lead to
\be
 {\cal A}(B \to D^{(*)}_s\pi ) &\sim &\int\!\! d x_1 d x_2 d x_3 b_1 d b_1 b_2 d
b_2 b_3 d b_3 \non && \cdot \mathrm{Tr} \left [ C(t) \Phi_B(x_1,b_1)
\Phi_{D^{(*)}_s}(x_2,b_2) \Phi_{\pi}(x_3, b_3) H(x_i, b_i, t) S_t(x_i)\,
e^{-S(t)} \right ], \quad\; \label{eq:a2}
\en
where $b_i$ is the
conjugate space coordinate of $k_{iT}$, and $t$ is the largest
energy scale in the function $H(x_i,b_i,t)$. The
last term $e^{-S(t)}$ in equation (\ref{eq:a2}) is the Sudakov form factor which suppresses
the soft dynamics effectively \cite{soft}.

\begin{figure}[t,b]
\vspace{-3cm} \centerline{\epsfxsize=16 cm \epsffile{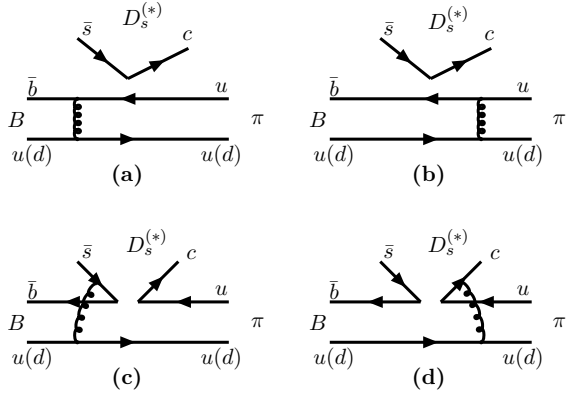}}
\vspace{-14cm} \caption{ Diagrams contributing to the decays $B\to D^{(*)}_s\pi$. }
 \label{fig1}
\end{figure}

 For the considered decays, the related weak effective
Hamiltonian $H_{eff}$ can be written as \cite{buras96}
\be
\label{eq:heff} {\cal H}_{eff} = \frac{G_{F}} {\sqrt{2}} \, V_{ub}^*
V_{cs}\left[ \left (C_1(\mu) O_1(\mu) + C_2(\mu) O_2(\mu)
\right)\right] ,
\en
where the four-quark operators are
\be
O_1=(\bar b_\alpha
u_\beta)_{V-A}(\bar c_\alpha s_\beta)_{V-A}  ,\quad O_2  = (\bar
b_\alpha u_\alpha)_{V-A}(\bar c_\alpha s_\alpha)_{V-A} ,
\en
with $\alpha, \beta$ being the color indexes, and $(\bar{q}_1q_2)_{V-A}=\bar q_1\gamma^\mu(1-\gamma^5)q_2$.
The Fermi constant $G_{F}=1.166 39\times 10^{-5} GeV^{-2}$ and
$C_{1,2}(\mu)$ are Wilson coefficients running with the renormalization scale $\mu$
. The leading order diagrams
contributing to the decays $B\to D^{(*)}_s\pi$ are drawn in
figure \ref{fig1} according to this effective Hamiltonian.

%===========================================================================
%                  Numerical results and discussions
%============================================================================

In the following, we will get the analytic formulas by calculating the hard part $H(t)$ at leading order.
Involving the meson wave functions, the amplitude for the factorizable tree emission diagrams Fig.1(a) and (b)
can be written as:
\be
F_e&=&8\pi C_Ff_{D^{(*)}_s}\int_0^1 dx_1 dx_3 \int_0^{\infty} b_1db_1\, b_3db_3\,
\Phi_B(x_1,b_1)\nonumber \\
& &\times
\left\{\left[(x_3+1)\phi^A_{\pi}(x_3)-r_\pi(2x_3-1)(\phi^P_{\pi}(x_3)+\phi^T_{\pi}(x_3))
\right]\right.\non &&\left.
\times E_e(t)h_e(x_1,x_3(1-r^2_{D^{(*)}_s}),b_1,b_3)S_t(x_3)\right.\non &&\left.
+2r_\pi\phi^P_{\pi}(x_3)E_e(t')h_e(x_3,x_1(1-r^2_{D^{(*)}_s}),b_3,b_1)S_t(x_1)\right\}\;, \label{fe}
\en
where $C_F=4/3$ is the group factor of $SU(3)_c$ gauge group, and the mass ratios $r_\pi=m_{0}^\pi/m_B, r_{D^{(*)}_s}
=m_{D^{(*)}_s}/m_B$. Here $f_{D^{(*)}_s}$ is the decay constant of $D^{(*)}_s$ meson, and $S_t(x)$ is the jet function \cite{PQCD}.
The factor evolving with the scale $t$ is given by:
\be
E_e(t)=\alpha_s(t)\exp[-S_B(t)-S_{\pi}(t)],
\en
where $S_B(t),S_{\pi}(t)$ are expressions for Sudakov form factors \cite{PQCD}. The hard function is written as
\be
h_e(x_1,x_2,b_1,b_2)&=&  K_{0}\left(\sqrt{x_1 x_2} m_B b_1\right)
\left[\theta(b_1-b_2)K_0\left(\sqrt{x_2} m_B
b_1\right)I_0\left(\sqrt{x_2} m_B b_2\right)\right. \non & &\;\left.
+\theta(b_2-b_1)K_0\left(\sqrt{x_2}  m_B b_2\right)
I_0\left(\sqrt{x_2}  m_B b_1\right)\right]. \label{he1}
\en
The hard scales $t^{(\prime)}$ in Eq.(\ref{fe}) are determined by
\be
t&=&max(\sqrt{x_3(1-r^2_{D^{(*)}_s})}m_B,1/b_1,1/b_3),\non  t'&=&max(\sqrt{x_1(1-r^2_{D^{(*)}_s})}m_B,1/b_1,1/b_3).
\label{scale1}
\en

For the nonfactorizable tree emission diagrams Fig.1(c) and (d), all three meson wave functions are
involved. The integraton of $b_3$ can be performed using $\delta$ function $\delta(b_3-b_2)$ and the result is
\be
M_e&=&-16\pi\sqrt{2N_c}C_F\int_0^1 dx_1dx_2dx_3 \int_0^{\infty} b_1 db_1\, b_2
db_2\,\Phi_B(x_1,b_1) \Phi_{D^{(*)}_s}(x_2) \non
&&\times \left\{\left[(x_2-1)\phi^A_{\pi}(x_3)+r_\pi x_3(\phi^P_{\pi}(x_3)-\phi^T_{\pi}(x_3))\right]
E_n(t)h^1_n(x_1,x_2,x_3,b_1,b_2)\right.\non &&\left.
+[(x_3+x_2)\phi^A_{\pi}(x_3)-r_\pi x_3(\phi^P_{\pi}(x_3)+\phi^T_{\pi}(x_3))E_n(t')h^2_n(x_1,x_2,x_3,b_1,b_2)]\right\},
\en
where the expressions for the evolution factor is $E_n=\alpha_s(t)\exp[-S(t)|_{b_3=b_1}]$ with the Sudakov exponent
$S=S_B+S_{D^{(*)}_s}+S_\pi$.

The hard functions $h^{i}_{n}, i=1,2$ in the amplitude are given as
\be
h^i_n&=&[\theta(b_1-b_2)K_0(Ab_1)I_0(Ab_2)+\theta(b_2-b_1)K_0(Ab_2)I_0(Ab_1)]\non &&
\non && \times\left(
\begin{matrix}
 \frac{\pi i}{2}\mathrm{H}_0(\sqrt{|G^2_i|} b_3), & \text{for}\quad G^2_i<0 \\
 \mathrm{K}_0(G_ib_2), &
 \text{for} \quad G^2_i>0
\end{matrix}\right),\label{cdhard}
\en
with the variables
\be
A^2&=&x_1x_3(1-r^2_{D^{(*)}_s})m^2_B,\non
G^2_1&=&(x_1+x_2)r^2_{D^{(*)}_s}-(1-x_1-x_2)x_3(1-r^2_{D^{(*)}_s})m^2_B,\non
G^2_2&=&(x_1-x_2)x_3(1-r^2_{D^{(*)}_s})m^2_B.
\en
The hard scales in Eq.(\ref{cdhard}) are given by
\be
t&=&max(Am_B,\sqrt{G^2_1}m_B, 1/b_1, 1/b_2), \non t'&=&max(Am_B,\sqrt{G^2_2}m_B, 1/b_1, 1/b_2).\label{scale}
\en

Then the total decay amplitude of $B\to D^{(*)}_s\pi$ decays can be written as
\be
{\cal A}(B\to D^{(*)}_s\pi)=
V_{ub}^* V_{cs}[F_{e}(C_2+\frac{C_1}{3})+M_{e}C_1]. \label{totalam}
\en
\section{Numerical results and discussions} \label{numer}

\begin{table}
\caption{Input parameters used in the numerical calculation\cite{pdg08,cleo}.}\label{para}
\begin{center}
\begin{tabular}{c |cc}
\hline \hline
 Masses &$m_{\pi}= 0.14\mbox{ GeV}$,   &$ m_0^{\pi}=1.3 \mbox{ GeV}$, \\
  & $ m_{D_s}=1.9685 \mbox{ GeV}$,&$m_{D^*_s}=2.1123 \mbox{ GeV}$,\\
  & $ m_B = 5.28 \mbox{ GeV}$,&$ m_{W}=80.4 \mbox{ GeV}$,\\
 \hline
  Decay constants &$f_B = 0.19 \mbox{ GeV}$,  & $ f_{\pi} = 0.13
 \mbox{ GeV}$,\\
 & $f_{D_s}=0.273 \mbox{ GeV}$,&$f_{D^*_s}=0.312 \mbox{ GeV}$,\\
 \hline
Lifetimes &$\tau_{B^\pm}=1.638\times 10^{-12}\mbox{ s}$, &
$\tau_{B^0}=1.530\times 10^{-12}\mbox{ s}$,\\
 \hline
$CKM$ &$V_{cb}=0.0412\pm0.0011$, & $V_{us}=0.2255\pm0.0019$.\\
\hline \hline
\end{tabular}
\end{center}
\end{table}

For the numerical calculation, we list the input parameters
in Table~\ref{para}.

For the $B$ meson wave function, we adopt the model
\be
\phi_B(x,b)
&=& N_B x^2(1-x)^2 \mathrm{exp} \left
 [ -\frac{M_B^2\ x^2}{2 \omega_{b}^2} -\frac{1}{2} (\omega_{b} b)^2\right],
 \label{phib}
\en
where $\omega_{b}$ is a free parameter and we take
$\omega_{b}=0.4\pm 0.04$ GeV in numerical calculations, and
$N_B=91.745$ is the normalization factor for $\omega_{b}=0.4$.

For $D^{(*)}_s$ meson, the distribution amplitude is taken as:
\be
\phi_{D^{(*)}_s}(x)=f_{D^{(*)}_s}\frac{1}{\sqrt 6}x(1-x)\left[1-a_{D^{(*)}_s}(1-2x)\right], \label{phids}
\en
with the Gegenbauer coefficients $a_{D_s}=0.3$ and $a_{D^{*}_s}=0.78$.
 The CLEO  and BarBar collaborations reported their work on the measurements of the decay constant of $D_s$
meson and obtained $f_{D_s}=274\pm13\pm7$ MeV \cite{cleo} and $283\pm17\pm7\pm14$ MeV \cite{barbar}, respectively. However, the decay constant of
the vector meson $D^{*}_s$ has not been directly measured in experiments so far. From the conclusions draw by the CLEO collaboration \cite{cleo}
, one can find that there exists a relation:
\be
\frac{f_{D^*_s}}{f_{D^*}}\approx\frac{f_{D_s}}{f_{D}}\approx\frac{f_{B_s}}{f_B}=[1.1,1.2],
\en
which is consistent with that from lattice simulation \cite{ukqcd} and the QCD sum rules calculations \cite{yuming}.
From table~\ref{para}, it is easy to see the value of the ratio $f_{D^*_s}/f_{D_s}$ is $1.14$ in our work. It is different from \cite{runhui},
where the relation between $f_{D^*_s}$ and $f_{D_s}$ derived from HQET
\be
\frac{f_{D^*_s}}{f_{D_s}}=\sqrt{\frac{m_{D_s}}{m_{D^*_s}}},
\en
was used. From this equation, one can get the value of $f_{D^*_s}$, which is smaller than that of $f_{D_s}$.

The twist-2 pion distribution
amplitude $\phi^{A}_{\pi}$, and the twist-3 ones $\phi^{P}_{\pi}$
and $\phi^{T}_{\pi}$ have been parametrized as
\begin{eqnarray}
 \ppi(x) &=&  \frac{f_\pi}{2\sqrt{2N_c} }
    6x (1-x)
    \left[1+a_1^{\pi}C^{3/2}_1(2x-1)+a^{\pi}_2C^{3/2}_2(2x-1)
    \right.\non && \left.+a^{\pi}_4C^{3/2}_4(2x-1)
  \right],\label{piw1}\\
 \ppip(x) &=&   \frac{f_\pi}{2\sqrt{2N_c} }
   \left[
   1+(30\eta_3-\frac{5}{2}\rho^2_{\pi})C^{1/2}_2(2x-1)-3\left\{\eta_3\omega_3+\frac{9}{20}\rho^2_\pi
   (1+6a^\pi_2)\right\}\right. \non && \left.\times C^{1/2}_4(2x-1)\right]  ,\\
 \ppit(x) &=&  \frac{f_\pi}{2\sqrt{2N_c} } (1-2x)
   \left[ 1+6(5\eta_3-\frac{1}{2}\eta_3\omega_3-\frac{7}{20}\rho^2_{\pi}-\frac{3}{5}\rho^2_\pi a_2^{\pi})
   (1-10x+10x^2)\right] ,\quad\quad\label{piw}
 \end{eqnarray}
 with the mass ratio $\rho_\pi=(m_u+m_d)/m_\pi=m_\pi/m_0^\pi$ and the
 Gegenbauer polynomials $C^{\nu}_n(t)$,
\be
C^{1/2}_2(t)&=&\frac{1}{2}(3t^2-1), \qquad C^{1/2}_4(t)=\frac{1}{8}(3-30t^2+35t^4), \\
C^{3/2}_1(t)&=&3t, \qquad C^{3/2}_2(t)=\frac{3}{2}(5t^2-1),\\
C^{3/2}_4(t)&=&\frac{15}{8}(1-14t^2+21t^4).\label{eq:c124}
\en
The Gegenbauer coefficients are given as \be
a^\pi_1=0 ,\quad a^\pi_2=0.115,\quad a^\pi_4=-0.015. \en
The values of
other parameters are taken as {\cite{bf}} $\eta_3=0.015$ and
$\omega=-3.0$.

In the B-rest frame, the decay width of $B\to D^{(*)}_s\pi$ can be obtained by
\be
\Gamma=\frac{1}{32\pi}G_F^2m^7_B|{\cal A}|^2(1-r^2_{D^{(*)}_s}),
\en
where ${\cal A}$ is the total decay amplitude shown in Eq.(\ref{totalam}).

%%%%%%%%%%%%%%%%%%%%%%%%%%%%%%%%%%
\begin{table}
\caption{Branching ratios ($\times 10^{-5}$) for the decays $B^0\to D_s^+\pi^-, D_s^{*+}\pi^-$ and
$B^+\to D_s^+\pi^0, D_s^{*+}\pi^0$. The first theoretical error is from
the the B meson shape parameter $\omega_b$. The second error is from the higher order pQCD correction.
The third one is from the uncertainties of CKM matrix elements. }\label{brch}
\begin{center}
\begin{tabular}{c|c|c}
   \hline \hline
   Channel & This work  &Data  \\
   \hline
    $B^0\to D_s^+\pi^- $ &$1.85^{+0.36+0.41+0.10}_{-0.52-0.56-0.10}$&$1.53\pm 0.35$\\
    $B^+\to D_s^+\pi^0 $&$1.98^{+0.39+0.81+0.11}_{-0.56-0.31-0.11}$&$1.6\pm0.6$\\
    $B^0\to D_s^{*+}\pi^-$ &$2.59^{+0.45+0.70+0.15}_{-0.76-0.60-0.15}$&$3.0\pm0.7$\\
    $B^+\to D_s^{*+}\pi^0$ &$2.78^{+0.48+0.74+0.16}_{-0.82-0.65-0.16}$&$<27$\\
   \hline\hline
\end{tabular}
   \end{center}
\end{table}
%%%%%%%%%%%%%%%%%%%%%%%%%%%%%%%%%%%%%%%%%%

Using the wave functions as specified in the previous section and the input parameters
listed in this section, it is straightforward to calculate the
CP-averaged branching ratios for the considered decays, which are listed in Table~\ref{brch}.
The first error in these entries is caused by the B meson shape parameter $\omega_b=0.40\pm0.04$.
The second error is from the higher order pQCD correction: the choice of hard scales, defined in Eq.(\ref{scale1}) and Eq.(\ref{scale}),
which vary from $0.9t$ to $1.1t$. The third error is from the
uncertainties of the CKM matrix elements which are listed in table~\ref{para}.

\begin{figure}[t,b]
\begin{center}
\includegraphics[scale=0.65]{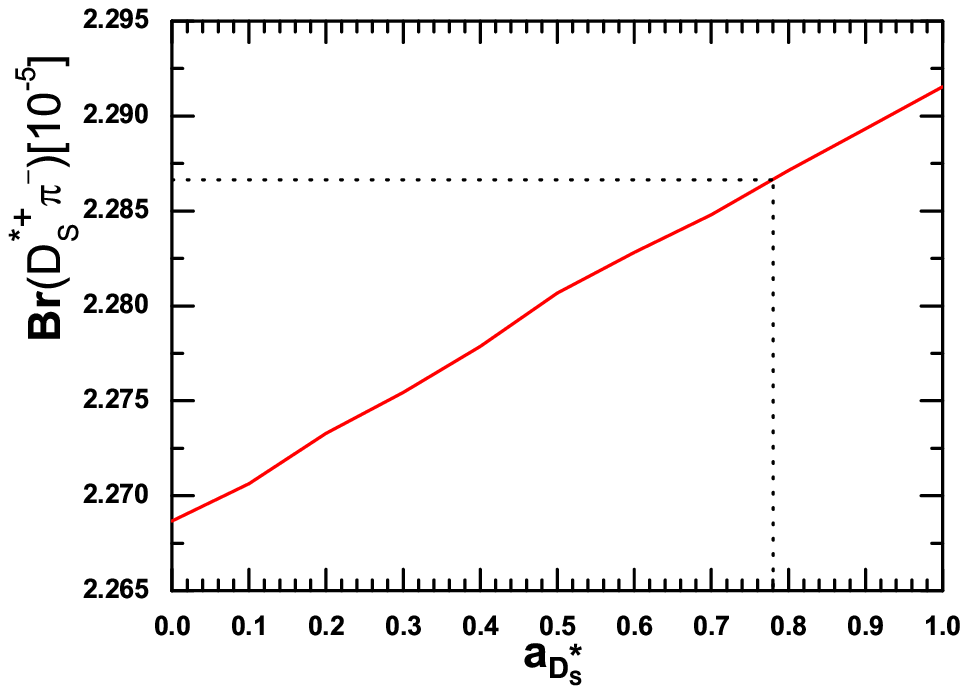}
\includegraphics[scale=0.65]{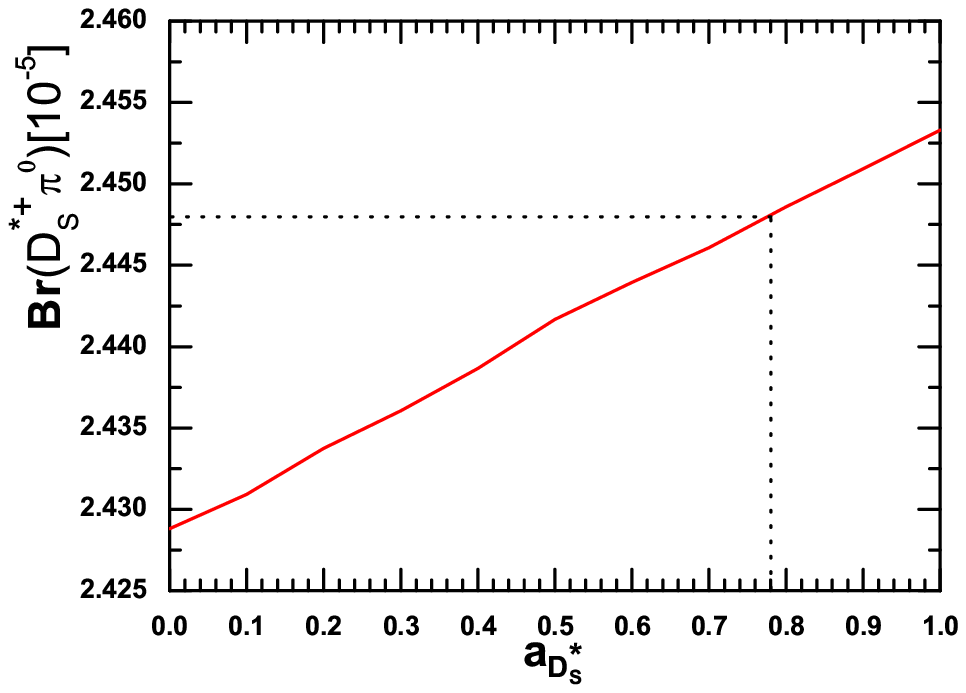}
\vspace{0.3cm} \caption{Branching
ratios (in units of $10^{-5}$) of  $B^0\to D_s^{*+}\pi^-$ and $B^+\to D_s^{*+}\pi^0$
decays as functions of Gegenbauer moment $a_{D^*_s}$ .}\label{dsstar}
\end{center}
\end{figure}

In previous calculations \cite{cdlv1,cdlv2}, the authors have considered that the value of the Gegenbauer moment $a_{D^*_s}$
was the same as that of $a_{D_s}$ and taken them as $0.3$. Here we take $a_{D^*_s}=0.78$, which is determined to fit
the requirement that $\phi_{D^*_s}(x)$, shown in Eq.(\ref{phids}), has
a maximum at $\bar{x}=\frac{m_{D_s}-m_c}{m_{D_s}}$. In Fig.~\ref{dsstar}, we plot that $a_{D^*_s}$ dependence of
the branching ratios of $B^0\to D^{+*}_s\pi^-$ and $B^+\to D^{+*}_s\pi^0$. One can find that the branching ratios
are not sensitive to the variations of $a_{D^{*}_s}$.

From the numerical results, we find that the non-factorizable contributions are very small and almost neglectable.
They are about $10\%$ of the factorizable ones in each decays. The main contributions come from
the factorizable amplitudes.

%===========================================================================
%                 Conclusion
%============================================================================

\section{Conclusion}\label{summary}

In this paper, we calculate the branching ratios of  decays $B^0\to D_s^+\pi^-, B^+\to
D_s^+\pi^0$, $B^0\to D_s^{*+}\pi^-$ and $ B^+\to D_s^{*+}\pi^0$
in the pQCD factorization approach. We find that:
\begin{itemize}
\item
The decays considered here have branching ratios about $10^2$ smaller than those of the $B\to D^{(*)}\pi$ decays, and they comes mainly from the relevant CKM matrix elements.
\item
From the numerical results shown in table~\ref{brch}, one can find that the pQCD predictions for these considered
decay channels are consistent with the measured values and currently available experimental upper limit.
\item
To determine decay constant of the vector meson $D^{*+}_s$ , the relation
\be
\frac{f_{D^*_s}}{f_{D^*}}\approx\frac{f_{D_s}}{f_{D}}\approx\frac{f_{B_s}}{f_B}
\en
is used. It indicates that the value of $f_{D^*_s}$ is larger than that of $f_{D^*}$, which is contrary to the conclusion derived from the relation
\be
\frac{f_{D^*_s}}{f_{D_s}}=\sqrt{\frac{m_{D_s}}{m_{D^*_s}}}\;.
\en
\item
In the numerical calculation, we take $a_{D^*_s}=0.78$, which is larger than the value given in the previous calculations. It is determined to fit the requirement that the wave function $\phi_{D^*_s}(x)$ has
a maximum at $\bar{x}=\frac{m_{D_s}-m_c}{m_{D_s}}$.

\end{itemize}

\section*{Acknowledgment}
Z.Q.~Zhang would like to thank C.D.~L\"u for fruitful discussions.

%%%%%%%%%%%%%%%%%%%%%%%%%%%%%%%%%%%%%%%%%%%%%%%%%%%%%%%%%%%%%%%%%%%%%%%%
%                               references
%%%%%%%%%%%%%%%%%%%%%%%%%%%%%%%%%%%%%%%%%%%%%%%%%%%%%%%%%%%%%%%%%%%%%%%%

\end{document}